\documentclass[a4paper,twoside,12pt]{article}

\usepackage{adm}

\begin{document}

\title[Приложение блочного метода Коркина--Золотарева в MIMO---декодировании]
      {Приложение блочного метода Коркина--Золотарева в MIMO---декодировании}%
      {Приложение блочного метода Коркина--Золотарева в MIMO---декодировании}
      {Application of lattice Reduction block Korkin--Zolotarev method to MIMO---decoding}
\author{В.\,С.~Усатюк}%
       {Усатюк~В.\,С.}
       {Usatyuk~V.\,S.}
\email{L@Lcrypto.com} 
\organization{}
\udk{511.9}
\maketitle

\begin{abstract}
 В работе была показана возможность применения блочного метода Коркина"=Золотарева с целью улучшения области принятия решения при декодировании сигналов в MIMO"=канале. Получена верхняя оценка точности декодирования по сравнению с методом максимального правдоподобия, в случае приведения базиса блочным методом Коркина"=Золотарева, при детектировании сигнала методом последовательного подавления помех.
\end{abstract}

Пусть у нас имеется MIMO"=система передачи, состоящая из $n$"---передающих и $m$"---принимающих антенн, которая осуществляет передачу $n$"---$q$"---ичных символов, $x\in C^{n}.$ Модель работы такой системы описывается уравнением:
\[y=Bx+\varepsilon ,\] 
\begin{flushright}
где $y$-сигнал полученный приемником $B\in C^{n\times m} $, $\varepsilon _{}^{} $"---гауссовый шум.
\end{flushright}

Детектирование исходного сигнала в MIMO методом максимального правдоподобия (ML) эквивалентно решению уравнения:

\[x_{ML} =\arg \mathop{\min }\limits_{x\in W} \left\| y-Bx\right\| ^{2}, \]
\begin{flushright}
 где $W$"---множество кодовых слов.
\end{flushright}

 В силу экспоненциальной сложности метода ML"---детектирования на практике применяются другие алгоритмы: обнуления (ZF), минимального среднеквадратичного отклонения (MMSE), последовательного подавления помех (SIC), см. \cite{YJWC2010}. 
 
 \begin{center}
\includegraphics[bb=0mm 0mm 209mm 297mm, width=53.9mm, height=50.9mm, viewport=4mm 4mm 205mm 293mm]{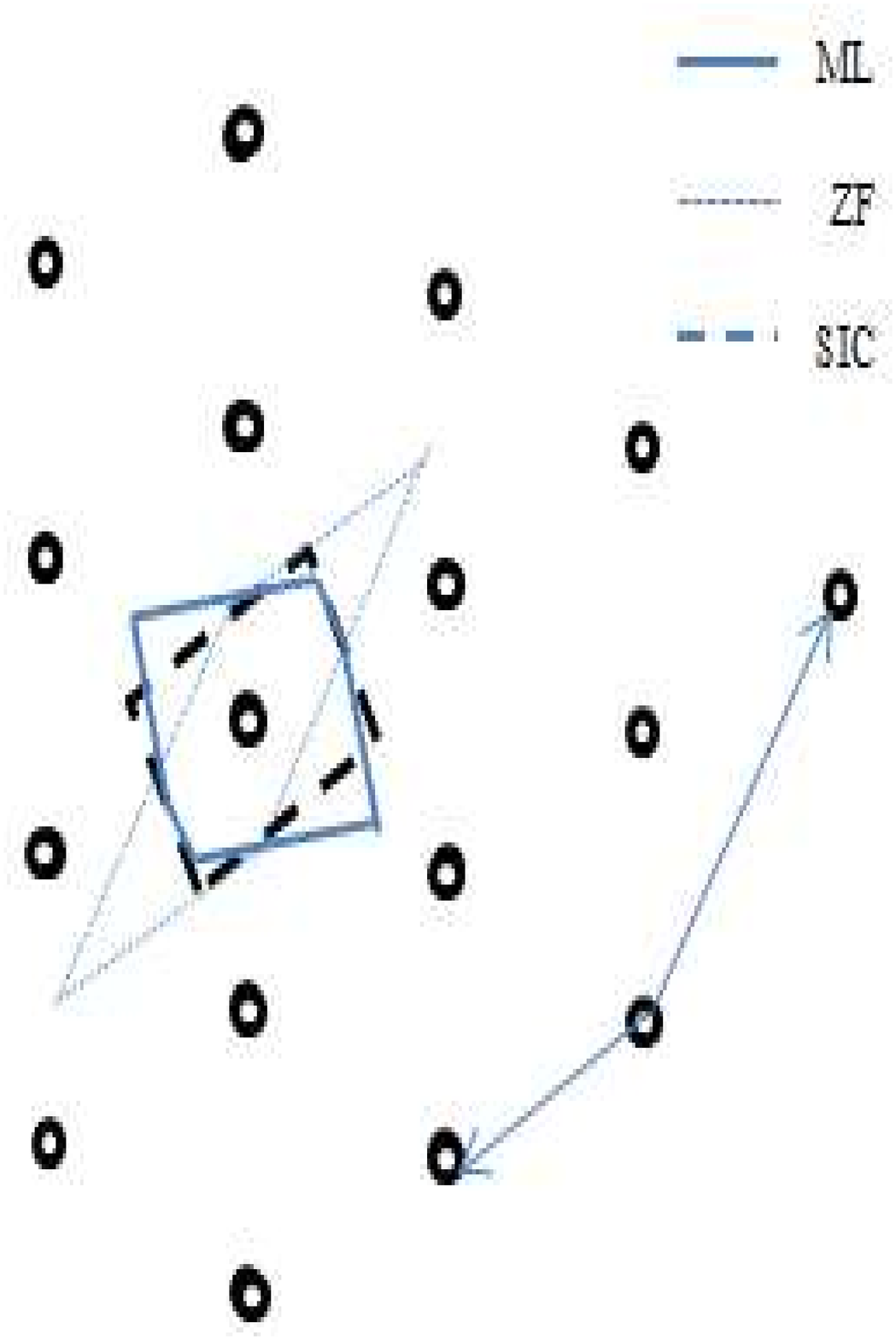}
 \end{center}
\begin{center} Рис. 1. Области принятия решения алгоритмами детектирования для приведенного базиса двухмерной решетки 
\end{center}

 Применение предварительного приведения канальной матрицы методами геометрии чисел позволяет получить хорошо обусловленную матрицу, тем самым улучшая область принятия решения (рис. 1), что в сочетании с вышеозначенными методами детектирования приводит к значительному энергетическому выигрышу, см. \cite{WSM2011}.

В работе \cite{CL2011}, были получены верхние оценки для детектирований обнулением и последовательным подавлением помех в сочетании с алгоритмами приведения канальной матрицы по Ленстра"=Ленстра"=Ловасу \cite{LLL1982} и Коркину"=Золотареву \cite{LLS2007} . Обобщим полученные оценки, в случае применения детектирования последовательным подавлением помех,  используя блочный метод Коркина"=Золотарева, см. \cite{S1994}.

\begin{Theorem} Пусть канальная матрица $B$ приведена блочным методом Коркина"=Золотарева, с размером блока $2\le \beta \le m$, тогда, в случае детектирования сигнала методом последовательного подавления помех, точность декодирования по сравнению с методом максимального правдоподобия не превосходит величину:

\[\rho _{SIC} =\gamma _{\beta }^{4\frac{m-1}{\beta -1} } \frac{m+3}{4} ,\] 

\begin{flushright}
где $\gamma _{\beta } $"---константа Эрмита, $m$"---ранг решетки.
\end{flushright}
\end{Theorem}

\begin{Proof}

По определению, точность декодирования по сравнению с методом максимального правдоподобия в случае предварительного приведения базиса и детектирования методом последовательного подавления помех, см.  \cite{CL2011}:

\begin{equation}
\label{formula}
\rho _{i,SIC} =\mathop{\sup }\limits_{B_{reduced} } \frac{\lambda ^{2} (L)}{\left\| b_{i}^{\bot } \right\| ^{2} } ,  
\end{equation}

где  $\lambda ^{2} (L)$-длина кратчайшего вектора в решетке, $\left\| b_{i}^{\bot } \right\| ^{2} $"---Евклидова норма ортогональных векторов в приведенном по Коркину"=Золотареву базисе.

В работе \cite{S1994}  были получены оценки для базиса решеток приведенного по Коркину"=Золотареву: 

\begin{center}
$\left\| b_{i} \right\| ^{2} \lambda _{i} (L)^{-2} \le \gamma _{\beta }^{2\frac{m-1}{\beta -1} } \frac{i+3}{4} ,\left\| b_{i}^{\bot } \right\| ^{2} \lambda _{i} (L)^{-2} \ge \gamma _{\beta }^{-2\frac{i-1}{\beta -1} } ,$ при $i=1,2...,m,$
\end{center}

где $\lambda _{i} (L)$- i-соответствующий минимум в решетке $L$, $\left\| b_{i} \right\| ^{2} $"---Евклидова норма векторов в приведенном по Коркину"=Золотареву базисе.

Поделив неравенства, получим $\left\| b_{i} \right\| ^{2} \le \gamma _{\beta }^{2\frac{m+i-2}{\beta -1} } \frac{i+3}{4} \left\| b_{i}^{\bot } \right\| ^{2} $. Подставив полученное неравенство в \eqref{formula}, получим:

\[\rho _{i,SIC} =\mathop{\sup }\limits_{B_{reduced} } \frac{\lambda ^{2} (L)}{\left\| b_{i}^{\bot } \right\| ^{2} } \le \gamma _{\beta }^{2\frac{m+i-2}{\beta -1} } \frac{i+3}{4} \left\| b_{i}^{\bot } \right\| ^{2} \mathop{\sup }\limits_{B_{reduced} } \frac{\lambda ^{2} (L)}{\left\| b_{i} \right\| ^{2} } .\] 

Из тривиальных соображений $\left\| b_{i}^{\bot } \right\| ^{2} \ge \lambda ^{2} (L)$. Откуда получим искомую оценку:

\[\rho _{SIC} \le \gamma _{\beta }^{4\frac{m-1}{\beta -1} } \frac{m+3}{4} .\] 
\end{Proof}

\enabstract{
This article present a application of Block Korkin---Zolotarev lattice reduction method for Lattice Reduction---Aided decoding under MIMO---channel. We give a upper bound estimate on the lattice reduced by block Korkin---Zolotarev method (BKZ) for different value of the block size and detecting by SIC.
\protect\enkeywords{lattices, MIMO, block Korkin---Zolotarev, BKZ.}
}

\begin{authors}
    \item{Усатюк Василий Станиславович}{}{L @Lcrypto.com}
\end{authors}


\begin{thebibliography}{1}

\bibitem{YJWC2010}
        \BibAuthor{Yong\;S.\,C.,Jaekwon\;K.,Won\;Y.\,Y. and etc}
        Wireless Communications with Matlab. 2010 Wiley-IEEE Press, 544 p. 


\bibitem{WSM2011}
        \BibAuthor{Wubber\;D.\,C., Seethaler\;J., Matz\;G.}
        \BibTitle{Lattice reduction}~// 
        Signal Processing Magazine. IEEE. 2011. V. 28. \No\,3. pp.\,70--91. 

\bibitem{CL2011}
        \BibAuthor{Cong\;L.}
        \BibTitle{On the Proximity Factors of Lattice Reduction-Aided decoding.}~//
		Signal Processing, IEEE Transactions on. 2011. V. 59. \No\,6. pp.\,2795--2808.


\bibitem{LLL1982}
        \BibAuthor{Lenstra\;A.\,C., Lenstra\;H., Lovasz\;L.}
        \BibTitle{ Factoring polynomials with rational coefficients.}~//
		 Math. Ann. 1982.  V. 261. \No\,4. pp.\,515--534.	       


\bibitem{LLS2007}
        \BibAuthor{Lagarias\;J.\,C., Lenstra\;H.\,W. Jr., Schnorr\;C.\,P.}
        \BibTitle{Korkin-Zolotarev bases and successive minima of a lattice and its reciprocal lattice.}~//
		Combinatorica. 1990. V. 10. \No\,4.  pp.\,333--348.
		
		
\bibitem{S1994}
        \BibAuthor{Schnorr\;C.\,P.}
        \BibTitle{Block reduced lattice bases and successive minima.}~//
		Combinatorics, probability and computing. 1994. V. 3.  pp.\, 507--522.
		
		
\end{thebibliography}
\end{document}